\documentclass[epj,nopacs]{svjour}
 \usepackage{amsfonts}
 \usepackage{amssymb,amsxtra}
 \usepackage{amsmath,exscale}
 \usepackage{latexsym}
 \usepackage{graphicx,wrapfig,subfigure,epsfig,here}
 \usepackage{cite}
\newcommand{\beq}{\begin{equation}}
\newcommand{\eeq}{\end{equation}}
\newcommand{\bea}{\begin{eqnarray}}
\newcommand{\eea}{\end{eqnarray}}
\newcommand{\bdm}{\begin{displaymath}}
\newcommand{\edm}{\end{displaymath}}

\newcommand{\nn}{\nonumber\\}

\newcommand{\Msig}{M_{\sigma}}

\newcommand{\e}{\epsilon}
\newcommand{\mufactor}{4\pi\mu^2}

\newcommand{\lns}{\ln\left(-\frac{s}{\mufactor}\right)}

\newcommand{\mael}[3]{\langle#1|#2|#3\rangle}
\newcommand{\abs}[1]{\left| #1 \right|}
\newcommand{\LN}[1]{\ln \left(#1 \right)}
\newcommand{\D}{{\cal D}}
\renewcommand{\d}[1]{\frac{\partial}{\partial #1}}

\newcommand{\ged}{\end{document}}

\begin{document}
\title{A renormalizable effective theory for leading logarithms in ChPT}
\author{M.~Bissegger\thanks{bissegg@itp.unibe.ch} \and A.~Fuhrer\thanks{afuhrer@itp.unibe.ch}}
%
%
\institute{Institute for Theoretical Physics, University of Bern,
  Sidlerstr. 5, CH-3012 Bern, Switzerland}
\date{Received: date / Revised version: date}
%
\abstract{
We argue that the linear sigma model at small external momenta is an effective theory for the leading
logarithms of chiral perturbation theory. Based on
this assumption an attempt is made to sum these leading logarithms using
the standard renormalization group techniques, which are valid in renormalizable
quantum field theories.
%
} 
\maketitle

\section{Introduction}
\label{sec:intro}
The theory of the strong interaction, Quantum Chromo Dynamics, 
does not allow the application of the powerful perturbation techniques in the low--energy
region. It is however possible to use the symmetries of QCD to construct an
effective theory in the low--energy region, chiral perturbation theory (ChPT)
\cite{Weinberg, GL1983, GL1985}, which allows a systematic perturbative
expansion of Green functions in powers of external  momenta and quark masses.\newline
In actual calculations in ChPT one expects the dominant contribution to stem
from the leading effective Lagrangian, which generates the leading chiral
logarithm. Even if the latter do not always dominate, it would be very interesting to know the leading chiral logarithms
to every order in the perturbative expansion, and to sum them up.\newline
In a recent publication \cite{BF}, we presented a procedure which allows the calculation
of leading logarithms of certain Green functions in the chiral limit rather
easily. In the present article, we address the question whether it is
possible to sum up these leading logarithms to all orders. \newline 
In a given renormalizable quantum field theory,  resummation of logarithms is based on 
the renormalization group equations (RGE). 
However, chiral perturbation theory is not renormalizable,
 and the structure of the RGE is therefore 
more involved \cite{colangelobuechler}.
In order to avoid the problems introduced 
by the nonrenormalizable nature of chiral perturbation theory, 
we consider a theory which {\it is
renormalizable} and reproduces the leading logarithms of chiral perturbation
theory. It is then natural to expect that the summation 
 of logarithms in this renormalizable 
theory can be performed by use of the RGE.\newline
It is known since long that the tree--level graphs of the 
linear sigma model reproduce, at small
momenta, the results of current algebra. In the modern language, this means
that they agree with the tree--level graphs of ChPT.
It was shown in Ref.~\cite{GL1983}, that this persists 
at one--loop order: provided that the
low--energy couplings in the chiral Lagrangian are properly adapted, Green
functions, evaluated in the linear sigma model at one--loop order, agree with
the result of ChPT at order $p^4$, see also \cite{Bettinelli}.
This shows that the linear sigma model
 is  a promising candidate for a renormalizable  theory
which generates the leading logarithms in ChPT.
\newline
To carry the comparison between ChPT and the linear sigma model to higher
orders in the momentum expansion, we consider
the  correlator of two scalar quark currents,
\bea\label{eq:corr}
H(s) &=&
i\int dx e^{ipx}\langle 0|T S^0(x)S^0(0)|0 
\rangle,\nn 
S^0&=&\bar uu+\bar dd\,;\, s=p^2\,,
\eea
in the chiral limit $m_u=m_d=0$. Its leading chiral logarithms have been 
worked out in ChPT to five--loop accuracy in \cite{BF}. This article is
devoted to an analysis of this correlator in the framework 
of the linear sigma model, addressing the questions just raised: Does the
linear sigma model reproduce these logarithms, and if yes, 
can they be summed?

The structure of the article is as follows: 
In section \ref{sec:llchpt}, we recall the structure of the leading
logarithms of $H(s)$ in chiral perturbation theory in the chiral limit.
 In section
\ref{sec:llsm}, we calculate the leading logarithms of the scalar two--point
function -- which corresponds to the quantity $H(s)$ -- in the 
linear sigma model, and show in section \ref{sec:comp}  that this theory 
 reproduces the leading logarithms of chiral 
perturbation theory up to and including two loops in this case.
 For this reason, we believe that the linear sigma model indeed is a
 renormalizable effective theory to calculate the leading 
logarithms in ChPT.
 In section \ref{sec:RG}, we consider the summation of leading 
logarithmic singularities in both,  the
symmetric as well as  in the spontaneously broken phase of the linear
sigma model. 
In the following section 
\ref{sec:twopoint},  we apply this 
technique to the scalar two--point function in the spontaneously 
broken phase. We are able to sum up a certain class of 
logarithmic terms, 
 and explain why an explicit summation of all leading logarithms is not 
possible with this technique. Finally, section 
\ref{sec:conc} contains a summary and concluding remarks. The appendices
contain several technical aspects of our investigation: 
In appendix A, we present expressions for triangle graphs, whereas 
the two--loop diagrams needed in the calculation of
the two--point function are displayed and discussed in appendix B. A dispersive
calculation used as a check on certain two--loop diagrams is presented in
appendix C, and scale dependent logarithms are summed up in appendix D.

\section{Leading logarithms in ChPT}\label{sec:llchpt}

In the chiral limit, the low--energy expansion of the 
scalar correlator can be written as
\bea
H(s) &=& \frac{B^2}{16\pi^2} \{P_0(s,\bar\mu)+P_1(s,\bar\mu)L+P_2(s,\bar\mu)L^2 + \cdots\}, \nn
L &=& \LN{-\frac{s}{\bar\mu^2}},\label{eq:H(s)}
\eea
where  $P_i$ are polynomials in $N = s/(16 \pi^2 F^2)$. The quantities
$B,F$ are the two low energy constants (LECs) at leading order in the chiral
expansion \cite{GL1983}, and the running scale of ChPT 
is denoted  by $\bar \mu$. The {\it leading terms} $\bar P_i$ of the
polynomials $P_i$ -- which are the coefficients of the leading logarithms --
are known up to five loops  \cite{BF},
\begin{align}
\bar P_0 &= 0, &\bar P_1 &= -6, &\bar P_2 &= 6N,\nn
\bar P_3 &= -\frac{61}{9}N^2, &\bar P_4 &= \frac{68}{9}N^3, 
&\bar P_5 &= -\frac{140347}{16200}N^4.
\end{align}
The full polynomials $P_i$ differ from $\bar P_i$ by terms of  
order $s^i$ and higher.

\section{Chiral logarithms in the linear sigma model}\label{sec:llsm}

We first introduce our notation of the linear sigma model and work out the
quantity in the linear sigma model which corresponds to the scalar correlator
$H(s)$. Then we calculate the two--loop leading logarithm of this quantity in
the linear sigma model.

\subsection{Notation}
The Lagrangian of the $O(4)$ linear sigma model 
coupled to  external scalar sources reads
\bea
\mathcal{L} &=&
\frac{1}{2}\partial_{\mu}\varphi^a\partial^{\mu}\varphi^a+\frac{m^2}{2}\varphi^a\varphi^a-\frac{g}{4}(\varphi^a
\varphi^a)^2+j^a\varphi^a,\nn
a &=& 0,...,3.\label{LLsigmaM}
\eea If $m^2 > 0$, the $O(4)$ symmetry is spontaneously broken down to
$O(3)$, leading to three Goldstone bosons. In order to expand around
the ground state $\varphi_G = (v,\vec{0})$ of the spontaneously broken theory,
one rewrites the Lagrangian with the shifted fields $\varphi =
(\phi+v,\vec{\pi})$ and the massless Goldstone bosons $\pi^a$ and the massive field
$\phi$ become visible in the Lagrangian,
\begin{eqnarray}
\mathcal{L} &=& \frac{1}{2}\left(\partial_{\mu} \phi \partial^{\mu} \phi +
  \partial_{\mu} \pi^a \partial^{\mu} \pi^a
\right)-\frac{1}{2}\left(3gv^2-m^2\right)\phi^2 +\nn
&& v K \phi - gv \phi^3 - \frac{g}{4}\phi^4 -\frac{g}{4} (\pi^a\pi^a)^2 +
\frac{1}{2} K \pi^a \pi^a\nn
&& -gv \phi \pi^a \pi^a -\frac{g}{2} \phi^2 \pi^a \pi^a + j^0 \phi +j^a \pi^a,\nn
K &=& m^2 - g v^2.\label{LLsigmaMbroken}
\end{eqnarray} To every order of the calculation, one has to determine $v$
such that the vacuum expectation value vanishes, $$\mael{0}{\phi(x)}{0} =
0.$$ 
To one--loop, the parameters have to be renormalised in the following way:
\begin{equation}
\begin{array}{ll}
\displaystyle g = \mu^{4-d} g_r \Big[1-24g_r \lambda\Big], &m^2 = m_r^2\Big[1-12 g_r \lambda\Big],\nn
\displaystyle \varphi = Z^{\frac{1}{2}}\varphi_R, &Z = 1+O(g_r^2),\nn
\multicolumn{2}{l}{\displaystyle
\lambda = -\frac{1}{32 \pi^2} \left( \frac{1}{\e}+\Gamma'(1)+\ln(4
  \pi)+1\right),}\nn
\displaystyle d = 4-2\e.\label{eq:g}
\end{array}
\end{equation}
For the vacuum expectation value $v$, one obtains
\begin{align}\label{eq:VEV}
v &= v_0\bigg[1-\frac{3 g_r}{16\pi^2}\ln\left(\frac{2 m_r^2}{\mu^2}
\right)+O(g_r^2)\bigg],\nn
v_0 &= \mu^{-\e}\frac{m_r}{\sqrt{g_r}}.
\end{align}

\subsection{Correspondence of the linear sigma model to chiral perturbation theory}

As shown in \cite{GL1983}, the generating functionals of the linear sigma
model (equipped with additional external fields) in the heavy mass limit and
chiral perturbation agree
at first nonleading order, provided the low--energy constants of chiral
perturbation theory are pertinent functions of the parameters of the linear sigma
model.\newline
We stick to our example, the scalar two--point function, and identify the corresponding
quantity in the linear sigma model. The external field $\chi^a$ -- which
couples to the quark condensate -- finds its counterpart in the external
scalar source $j^a\,\,$\footnote{This identity only holds up to a finite
  renormalization factor which is a polynomial in the renormalized coupling
  constant $g_r$. However, this factor does not affect the leading
  logarithms.}. Therefore, the counterpart of $H(s)$ is the renormalized scalar two--point function
\bea \label{eq:stf}
G^{(2,\vec{0})}_R(s) = iZ \int d^4x e^{ipx}\mael{0}{T\phi(x)\phi(0)}{0}, \, s = p^2,
\eea
for small external momenta $s$.

\subsection{Leading logarithm to two loops}

We calculate the leading logarithms to one and two loops in the quantity $G^{(2,\vec{0})}_R(s)$.
In the following, we only quote the result and relegate the
description of the calculation and the individual loop contributions to appendix \ref{app:twoloop}.\newline
It is evident that $G^{(2,\vec{0})}_R(s)$ for small external momenta has the structure
\bea
G^{(2,\vec{0})}_R(s,g_r,m_r^2,\mu) &=& \frac{1}{2
  m_r^2}\Big[c^{(0)}(s,m_r^2,\mu)+c^{(1)}(s,m_r^2,\mu) g_r \nn
&&+ c^{(2)}(s,m_r^2,\mu) g_r^2 +  O(g_r^3)\Big].\label{eq:twopointlsm}
\eea
We now decompose the coefficients $c^{(i)}$ and indicate all logarithms which
are possible at the corresponding order in $g_r$:
\bea 
c^{(0)} &=& a^{(0)}_{0,0}\,,\nn
c^{(1)} &=& a^{(1)}_{1,0}\,L_s+a^{(1)}_{0,1}\,L_m+a^{(1)}_{0,0}\,,\nn
c^{(2)} &=& a^{(2)}_{2,0}\,L_s^2+a^{(2)}_{1,1}\,L_s L_m+ a^{(2)}_{0,2}\, L_m^2
+a^{(2)}_{1,0}\,L_s \nn
&&+ a^{(2)}_{0,1}\, L_m + a^{(2)}_{0,0},\nn
&\vdots&
\eea
\vspace{-1.0cm}
\begin{align}\label{eq:deltaF}
L_s &= \LN{-\frac{s}{\mu^2}},  &L_m = &\LN{\frac{2m_r^2}{\mu^2}}.
\end{align}
The coefficients $a^{(k)}_{l,m}$ are polynomials in $s/m_r^2$.
The indices of a coefficient $X_{k,l}^{(N,t)}$ always have the same meaning in
the following: The lower indices $k$ and $l$ indicate the power of the
momentum and mass logarithms $L_s$ and $L_m$, respectively. The upper indices
$N$ and (if present) $t$ stand for the order of the coupling constant $g_r$
and the power of $s/m_r^2$, respectively.\newline
In general, the coefficient $c^{(k)}$ can be written as a double sum
\beq
c^{(k)} =
\sum_{n=0}^{k}\sum_{l=0}^{k-n}a^{(k)}_{l,k-n-l}L_s^lL_m^{k-n-l}\,\,.
\eeq
The coefficients $a^{(k)}_{l,m}$ are given by
\begin{align}
a_{0,0}^{(0)}&=1+\frac{1}{2}\frac{s}{m_r^2}+\cdots,\nn
a_{0,0}^{(1)}&=-\frac{3}{8\pi^2}-\frac{21}{64\pi^2}\frac{s}{m_r^2}+\cdots,\nn
a_{1,0}^{(1)}&=-\frac{3}{16\pi^2}-\frac{3}{16 \pi^2}\frac{s}{m_r^2}+\cdots,\nn
a_{0,1}^{(1)}&=-\frac{3}{16\pi^2}-\frac{3}{16 \pi^2}\frac{s}{m_r^2}+\cdots,\nn
a_{2,0}^{(2)}&= \frac{3}{256 \pi^4}\frac{s}{m_r^2}+\cdots, \nn
a_{1,1}^{(2)} &= -\frac{9}{128 \pi^4}-\frac{3}{128 \pi^4}\frac{s}{m_r^2}+\cdots,\nn
a^{(2)}_{1,0} &=\frac{3}{128 \pi^4}+\frac{21}{256 \pi^4} \frac{s}{m_r^2}+\cdots.
\label{eq:2pointcoef}
\end{align} 
As an independent check of our loop calculation, we worked out the
discontinuity of $G^{(2,\vec{0})}_R(s)$,
\bea\label{eq:disclinsig}
G^{(2,\vec{0})}_R(s+i\epsilon)-G^{(2,\vec{0})}_R(s-i\epsilon)=2i\pi\rho(s)
\eea
and compare with the discontinuity obtained from the optical theorem. The two
expressions agree at the order considered. We refer to appendix \ref{app:dispersive} for further details.

\section{Linear sigma model versus ChPT}\label{sec:comp}

The translation rules provided in \cite{GL1983} are
\begin{eqnarray}
   B&=&\frac{\sqrt{g_r}}{2m_r}\left(1+\frac{1}{16\pi^2}\left(3L_m-1\right)g_r + O(g_r^2)\right),\nn
   F^2&=&\frac{m_r^2}{g_r}\left(1-\frac{1}{16\pi^2}(6L_m-1)g_r +
     O(g_r^2)\right).
\end{eqnarray}
Note that the coupling constant $g_r$ differs from the one introduced
in \cite{GL1983} by a term of order $g_r^2$. The higher--order corrections to the above relations do not
affect the coefficients of the leading logarithms $a^{(N)}_{N,0}$.\newline
Translating with the above rules the coefficient of the one-- and
two--loop leading logarithms of the scalar correlator in
Eq.~(\ref{eq:H(s)}) leads to
\begin{align}
\frac{B^2 \bar{P}_1}{16\pi^2}&=\frac{1}{2
  m_r^2}\left(-\frac{3}{16\pi^2} g_r
  +\frac{3}{128\pi^4}(1-3L_m) g_r^2 + O(g_r^3) \right)\nn
&=\frac{1}{2 m_r^2} \left( a_{1,0}^{(1,0)} g_r + (a_{1,0}^{(2,0)} +
  a_{1,1}^{(2,0)} L_m) g_r^2 + O(g_r^3)\right),\nn
\frac{B^2\bar{P}_2}{16 \pi^2}&=\frac{1}{2 m_r^2}\cdot\frac{3}{256
  \pi^4}\frac{s}{m_r^2}g_r^2+O(g_r^3)\nn
&=\frac{1}{2m_r^2}a_{2,0}^{(2,1)}g_r^2+O(g_r^3).
\end{align} 
It is seen that they agree at the
order considered. We have checked that the coefficients $a_{0,0}^{(0,0)}$, $a_{0,0}^{(1,0)}$ and
  $a_{0,1}^{(1,0)}$ agree as well. Therefore the one-- and two--loop leading logarithms of the linear sigma
model are the same as the one-- and two--loop leading logarithms in chiral
perturbation theory in this correlator.\newline
We take this result as strong evidence that the leading logarithms of both theories
agree to all orders in perturbation theory. Further support for this
conjecture is the fact that, as shown in Ref.~\cite{BF}, the
leading logarithms in the scalar two point function in ChPT are determined
by the tree--level amplitude. Stated differently, we believe that the
linear sigma model acts as a renomalizable effective field theory for
the leading logarithms in ChPT.\newline
In the remaining part of this article, we assume that our conjecture
is correct, and work out its consequences: summing leading logarithms
in the linear sigma model amounts to summing leading logarithms of the
pertinent quantities in ChPT. 

\section{Renormalization group analysis in the linear sigma model}\label{sec:RG}

In this section, we illustrate the summation of leading logarithms
with renormalization group techniques in the symmetric as well as in
the spontaneously broken phase and investigate the low--energy structure of the
correlator $G^{(2,{\bf 0})}_R(s)$.

\subsection{Symmetric phase}
Here we consider mass logarithms in the perturbative expansion
of the physical mass (i.e., the position of the pole in the two--point 
function) in the symmetric phase of the linear sigma model.
In particular, we recall how the leading, next-to-leading, etc. logarithms
can be explicitly summed up.\\  
First we recall the renormalization group equation (RGE) in the unbroken phase
of the linear sigma model for renormalized, Fourier transformed Green functions in four
space-time dimensions $G_R^{(\bold{n})}(p_i;g_r,m_r^2,\mu)$
\bea
\Big(\D&+&\sum_{k=1}^4 n_k\gamma\Big){G}_R^{(\bold{n})}(p_i)
= 0;\nn
\bold{n}&=&(n_1,n_2,n_3,n_4),\label{eq:RGEunbrokenphase}
\eea
where 
\bea\label{eq:phi4beta}
\D &=& \mu \d{\mu}+\beta \d{g_r}-m_r^2 \gamma_m \d{m_r^2},\nn
\beta &=& \mu \d{\mu}g_r = \sum_{k=2}^{\infty}\beta^{(k)}g_r^k=\frac{3}{2 \pi^2}g_r^2 +O(g_r^3),\nn
\gamma_m &=& -\frac{1}{m_r^2} \mu \d{\mu}m_r^2 = \sum_{k=1}^{\infty}\gamma_m^{(k)}g_r^k=-\frac{3}{4 \pi^2}g_r+O(g_r^2)  ,\nn
\gamma &=& \frac{1}{2} \beta \d{g_r} \log{Z} = O(g_r^2).
\eea
In the perturbative expansion, the physical mass has the structure
\begin{eqnarray}
m_{ph}^2 &=& m_r^2 \left(k^{(0)} + k^{(1)} g_r + k^{(2)} g_r^2 + \cdots\right),
\end{eqnarray}
where
\begin{eqnarray}
k^{(n)} &=& k_{n}^{(n)} L_{\phi^4}^n +  k^{(n)}_{n-1} L_{\phi^4}^{n-1} + \cdots +k^{(n)}_{0}\, ; \, k^{(0)}=1,\nn
L_{\phi^4}&=&\ln\left(\frac{m_r^2}{4\pi\mu^2}\right).
\end{eqnarray}
The leading logarithms are fully 
determined by the one--loop expressions $\beta^{(2)}$ and $\gamma_m^{(1)}$.
The proof of this statement (following the lines of \cite{ahmady}) starts from the observation
that the physical mass obeys the homogeneous RGE
\bea\label{eq:phimassunbroken}
\D m^2_{ph}&=&0.
\eea
Collecting the coefficients proportional to $g_r^NL_{\phi^4}^{N-1}$, which must
vanish individually, we find the recursion relation
\begin{align}
&-2 N k_{N}^{(N)} + \left\{(N-1)\beta^{(2)} - \gamma_m^{(1)}\right\} k_{N-1}^{(N-1)}
\label{eq:phi4recurs}
= 0;\nn
&N=1,2,\ldots .
\end{align}
It is seen that the one--loop expressions for the $\beta$- and $\gamma_m$-functions
suffice to determine the coefficients $k^{(N)}_{N}$. In order to sum the  logarithms, we introduce the
quantities
\bea
f_i(x)&=&\sum_{n=0}^\infty k^{(i+n)}_{n}x^n;\, x=g_rL_{\phi^4},\nn
m_{ph}^2&=&m_r^2\sum_{i=0}^\infty f_i(x)g_r^{i}.
\eea
\begin{figure}
\centering
\includegraphics[height=5cm]{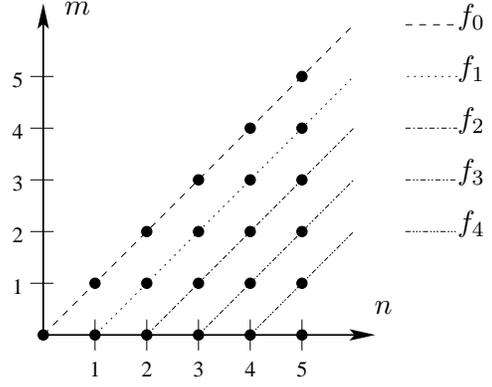}
\caption{Illustration of the structure of physical mass in the
  symmetric phase of the linear sigma model. The
  quantity $n$ represents the order in $g_r$, $m$ stands for the exponent
  of the logarithm $L_{\phi^4}$, the points represent coefficients
  $k^{(n)}_m$ and the different dotted lines stand for
  the connection between them.\label{fig:symmetricphaserelation}}
\end{figure}
The $f_i$ correspond to the sum of terms along the tilted lines
in Fig.~\ref{fig:symmetricphaserelation}; in particular, $f_0\, (f_1)$ denotes the sum of the 
leading (next-to-leading) logarithms. 
From the recursion relation (\ref{eq:phi4recurs}) it
follows that $f_0$ satisfies the differential equation
\bea
\left\{\left(2-x\beta^{(2)}\right)\frac{d}{dx}
+\gamma_m^{(1)}\right\}f_0(x)=0,
\eea
from where one has
\bea
f_0(x)&=&\left(1-\frac{\beta^{(2)}}{2}x\right)^{\frac{\gamma_m^{(1)}}{\beta^{(2)}}}=
\left(1-\frac{3}{4\pi^2}x\right)^{-\frac{1}{2}}. 
\eea
The next-to-leading logarithms can be summed up in an analogous fashion. 
It is easy to convince oneself that one needs a two--loop calculation
of the $\beta$- and $\gamma_m$-functions in this case.

\subsection{Spontaneously broken phase}
The derivation of the renormalization group equations 
for the linear sigma model in the spontaneously broken phase goes through exactly like in
the unbroken phase,
\begin{align}
&\left(\D+(k+\sum_{t=1}^3j_t)\gamma\right){G}^{(k,\bold{j})}_R(p_i)
= 0\label{RGE2};\nn
&\bold{j}=(j_1,j_2,j_3).
\end{align}
Here we have denoted the renormalized Fourier transformed Green function with $k (j)$ sigma
(pion) fields by $G_R^{(k,\bold{j})}$.
As in the symmetric phase, it is straightforward to sum up the leading
logarithms of quantities which depend only on two scales. We illustrate this statement with the vacuum expectation
value and the zero of the inverse sigma propagator. 

\subsubsection{Vacuum expectation value}
The vacuum expectation value of the sigma field fulfills
the inhomogeneous renormalization group equation
\beq
(\D+\gamma)v(g_r, m_r^2, \mu) = 0.
\eeq
The perturbative series of $v$ has the form
\bea
v &=& \frac{m_r}{\sqrt{g_r}}\left(v^{(0)}+v^{(1)}g_r+v^{(2)}g_r^2+\cdots\right),\nn
v^{(n)} &=& v^{(n)}_nL_m^n+v^{(n)}_{n-1}L_m^{n-1}+\cdots+v^{(n)}_0;\, v^{(0)}
= 1.
\eea
The recursion relation for the coefficients of the leading logarithms reads
\beq\label{eq:vacuumrecurs}
2 N v^{(N)}_{N}+\Bigg\{
\beta^{(2)}\left(\frac{3}{2}-N\right)+\frac{1}{2}\gamma_m^{(1)}\Bigg\} v^{(N-1)}_{N-1}
= 0.
\eeq Collecting  again the leading logarithms in a function $h_0(x)$ leads to the
differential equation

\bea
\Bigg\{
\frac{1}{2}\left(\beta^{(2)}+\gamma_m^{(1)}\right)+\left(2-\beta^{(2)}x\right)\frac{d}{dx}
\Bigg\}h_0(x) &=& 0,
\eea where $x = g_r L_m$, with the solution
\bea
h_0(x) &=&
\left(1-\frac{\beta^{(2)}}{2}x\right)^{\tfrac{\gamma_m^{(1)}+\beta^{(2)}}{2
    \beta^{(2)}}} = \left(1-\frac{3}{4 \pi^2}x\right)^{\tfrac{1}{4}}.
\eea

\subsubsection{The zero of the inverse sigma propagator}\label{sec:zeroinv}
Next we investigate the zero $\hat M$ of the inverse sigma propagator. 
We denote  by $\mbox{Re}(\hat{M})$ its real part, and find
\bea\label{eq:mass}
\mbox{Re}(\hat{M}) &=& 2m_r^2 \sum_{n=0}^{\infty}g_r^n \sum_{i=0}^{n} b^{(n)}_{i}L_m^i\nn
&=& 2m_r^2 \sum_{i=0}^\infty p_i(x) g_r^i,\nn
p_0(x) &=&
\left(1-\frac{\beta^{(2)}}{2}x\right)^{\tfrac{\gamma_m^{(1)}}{\beta^{(2)}}} =
\left(1-\frac{3}{4 \pi^2}x\right)^{-\tfrac{1}{2}}.
\eea 
Note that in the broken phase, the functions $p_i(x)$ are the {\it same} as in
the symmetric phase. Therefore, the coefficients of the mass logarithms in
$\mbox{Re}(\hat{M})$ and in the physical mass of the symmetric phase coincide
up to a factor of $2$.

\section{Summing leading logarithms?}
\label{sec:twopoint}
Here, we apply renormalization group techniques to the correlator
$G^{(2,\bf{ 0})}_R(s)$, written in the form Eq.~(\ref{eq:twopointlsm}), with
an attempt to sum the leading logarithms $a^{(N)}_{N,0}L_s^N$. To
start with, we insert the right hand side of
Eq.~(\ref{eq:twopointlsm}) into the RGE Eq.~(\ref{RGE2}). As the coefficients $a_{k,l}^{(n)}$ are analytic functions in $\tfrac{s}{m_r^2}$
they can be represented by a power series
\bea
a_{k,l}^{(n)}&=&\sum_{t=0}^{\infty}a_{k,l}^{(n,t)}\left(\frac{s}{m_r^2}\right)^t=a^{(n,0)}_{k,l}+a^{(n,1)}_{k,l}\frac{s}{m_r^2}+....
\eea
Analyticity demands the disappearance of the terms
proportional to $L_s^iL_m^j$ individually and leads to the recursion relations
for the leading momentum logarithms,
\begin{equation}
\begin{aligned}[b]
-2& N a^{(N,t)}_{N,0} - 2 a^{(N,t)}_{N-1,1}\\ &+ \left( (N-1) \beta^{(2)}
 + (1+t) \gamma_{m}^{(1)}\right)
a^{(N-1,t)}_{N-1,0}  =  0.\label{eq:recursionrelationbroken}
\end{aligned}
\end{equation}
\begin{figure}
  \centering
  \includegraphics[height=5cm]{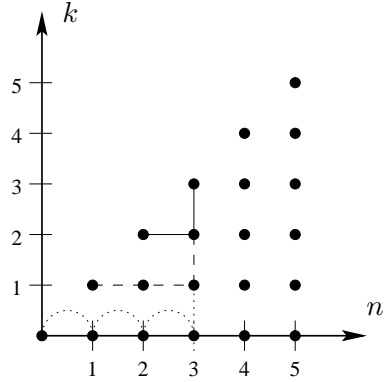}
  \caption{Illustration of the connections between the coefficients of the scalar two--point function in
    the spontaneously broken phase of the linear sigma model at order $g_r^3$. The
  quantity $n$ represents the order in $g_r$ and $k$ stands for the exponent
  of the logarithm $L_s$. Every type of line indicates recursion relations
  containing the connected coefficients. There is one such picture for every
  order in $\tfrac{s}{m_r^2}$. \label{fig:brokenphaserelation}}
\end{figure}
From this relation one concludes that in every order in
$s/m_r^2$ such an equation exists. This is
manifested by the index $t$. Furthermore this recursion relation connects the coefficient of the leading
logarithm at order $g_r^N$, $a_{N,0}^{(N,t)}$, with the coefficient of
the leading logarithm at order $g_r^{N-1}$, $a_{N-1,0}^{(N-1,t)}$, and
with the part of the coefficient of the next-to-leading logarithm at order
$g_r^N$ which is
proportional to one mass logarithm, $a_{N-1,1}^{(N,t)}$. In addition
only the one--loop results of the $\beta$- and
$\gamma_m$-function, $\beta^{(2)}$ and $\gamma_m^{(1)}$, appear in the
recursion relation.\\
Comparing with the previous recursion relations for the physical mass and for the vacuum expectation value, one finds that in these relations
only the coefficients of leading logarithms are involved. This
fact allows the summation of the leading logarithms. In
Eq.~(\ref{eq:recursionrelationbroken}), however,
the coefficients of the leading logarithms are no longer
connected directly. This is
illustrated in Fig.~\ref{fig:brokenphaserelation} by the dint of the
solid line. First one could expect that there still exist recursion
relations which allow a direct connection between the coefficients of
the leading logarithms. However, the dashed lines stand for recursion relations without leading logarithm
coefficients and demonstrate that this idea is not successful. Therefore the summation
of the leading logarithms fails, since the troublesome coefficient
$a_{N-1,1}^{(N,t)}$ is only determined by a $N$--loop calculation. \newline
As seen above, only coefficients with power $t$ in $s/m_r^2$
enter in the recursion relation. On the other hand the coefficient of the leading logarithm at
order $g_r^N$ is proportional to $(s/m_r^2)^{N-1}$, hence
this recursion relation cannot relate them.\newline
Therefore one cannot
determine the leading logarithm at order $g_r^N$ with the knowledge of
the leading logarithm at lower order. For this reason the summation fails.\\
The situation becomes clear by introducing a new scale $\rho$ and splitting up all
mass and momentum logarithms as 
\begin{align}
L_s &= \LN{-\frac{s}{\rho^2}}+L_{\mu}, &L_m &=
\LN{\frac{2m_r^2}{\rho^2}}+L_{\mu},\nn
L_{\mu} &= \LN{\frac{\rho^2}{\mu^2}}.
\end{align}
Therefore, all terms of the form $g_r^N L_s^k L_m^l$ with $k+l = N$ in the scalar two--point
function Eq.~(\ref{eq:twopointlsm}) generate a logarithm
$L_{\mu}^N$.  At a given order $g_r^N$, one is left with one
$\mu$-dependent logarithm with power $N$. The leading logarithms $L_{\mu}$ can
be summed to all orders, as we show explicitly in appendix
\ref{app:summation}. It is now obvious that only all explicit scale dependent
logarithms $L_\mu$ can be summed with the help of the RGE.\newline
In the case of the vacuum expectation value and the zero
of the inverse propagator, the coefficients of the logarithms $L_m$ and
$L_{\mu}$ are the same because there are only two scales involved. In the
presence of three scales, this is no longer true and a separation between
the leading momentum logarithms $L_s^N$ and other logarithms to the power $N$ like
$L_s^kL_m^l$ with $k+l = N$ is no longer possible with this technique.\newline\newline
Another access to the recursion relation is the solution of the
Callan-Symanzik equation which provides a relation between $n$--point functions with
momentum $p_i$ and the scaled momentum $p_i/\xi$. But the recursion
relations obtained in this way can be extracted from the ones worked out with
the RGE. Therefore the Callan-Symanzik equation does not contain new information.

\subsection{Linear sigma model with scale independent counterterms}

In chiral perturbation theory, the leading logarithms are in principle always accessible
with a one--loop calculation \cite{colangelobuechler}. One might hope to transfer this method to the
linear sigma model by using a formulation of the linear sigma
model with scale independent counterterms, analogously to chiral perturbation
theory. This formulation is discussed in \cite{piscattering}.
Studying the simplest case, we tried to calculate the one--loop leading
momentum logarithm of the scalar two--point function with the help of the tree--level
diagram containing the counterterm. One observes that only the sum of the
coefficients of the leading momentum and the leading mass logarithm
can be obtained in this manner. Therefore the statement is the same as with the
recursion relations in the previous subsection.

\section{Summary and conclusion}\label{sec:conc}

In this article, we investigate the structure of
 leading chiral logarithms  in the correlator of two scalar 
quark currents, Eq.~(\ref{eq:corr}). In particular, 
we determine this correlator
in the framework of the linear sigma model and compare the 
result with what is known from ChPT.

As a first step, we show that the leading logarithms agree  in the two
 theories at order $p^6$ in the low--energy expansion (two--loop order).
 To the best of our knowledge, this is a new result and
strongly suggests that the linear sigma model can be used as a 
{\it renormalizable} effective theory to calculate leading logarithms 
in $SU(2)\times SU(2)$ ChPT. The result also suggests that renormalization
group techniques can be used to sum these terms. 
For this reason, we investigate the RG equation in the linear sigma 
model and use it to sum up leading mass singularities 
e.g. in the vacuum expectation value of the sigma field.\newline
Applying the same technique to the scalar two--point 
function $G^{(2,\vec{0})}_R(s)$ -- which is the analogue of the correlator $H(s)$
in Eq.~(\ref{eq:corr}) -- allows one to
work out recursion relations between the coefficients of the leading
logarithms. We show that these recursion relations also contain subleading
terms,
 which are not accessible by the renormalization group. 
As a result of this, given the leading logarithm at order $g_r^N$, 
the recursion relations do not allow one to calculate the leading logarithm
 at order $g_r^{N+1}$.\newline
A summation of the explicit scale dependent leading logarithms is none\-theless always
possible. However, if there are more than two scales involved, a separation
between different types of leading logarithms like $\ln^N{(-s/\mu^2)}$  
and $\ln^N{(2m_r^2/\mu^2)}$, for example,
is not possible. Therefore, an independent summation of the leading momentum
logarithms fails, it is  only the sum of all coefficients 
of explicit scale dependent leading
logarithms which is accessible. 
In the special case of only two scales (for example
$\mu$ and $m_r$), the coefficients of the explicit scale dependent logarithms trivially agree with the
coefficients of the leading mass logarithms.\newline
To conclude, even if the linear sigma model represents an effective
renormalizable theory for the leading logarithms of chiral perturbation
theory, the summation of these leading logarithms 
by a straightforward use of the renormalization group seems not 
to be possible.

\section{Acknowledgments}

We would like to thank C.~Greub and all the members of the institute for informative discussions. Furthermore, we
are indebted to J.~Gasser for a careful reading of the manuscript and many useful comments. This work was supported in part by the Swiss National Science Foundation,
by RTN, BBW-Contract No. 01.0357, and EC-contract HPRN-CT2002-00311 (EURIDICE).

\begin{appendix}

\newcounter{zahler}
\renewcommand{\thesection}{\Alph{zahler}}
\renewcommand{\thesubsection}{\Alph{zahler}.\arabic{subsection}}
\renewcommand{\theequation}{\Alph{zahler}\arabic{equation}}

\setcounter{zahler}{0}

\newpage
\setcounter{equation}{0}
\addtocounter{zahler}{1}

\section{Triangle integrals}

Most of the one-- and two--loop integrals which are used in the loop calculations in appendix \ref{app:twoloop} and
\ref{app:dispersive} are provided in  Ref.~\cite{ScharfTausk}. However, triangle
integrals with one, two and three massless particles propagating in 
the loop are not considered there. This is the reason why we indicate the
results of these vertex functions here.
\bea
C^{(1)}(s, m^2) &=& \int\frac{d^dl}{i(2\pi)^d}\frac{1}{
  m^2-(l+k_1)^2}\frac{1}{m^2-(l-k_2)^2}\frac{1}{-l^2}\nn
&=& \frac{1}{16 \pi^2 m^2}\left(1+\frac{\tau}{12}  +O(\tau^2) \right),\nn
C^{(2)}(s, m^2)&=&\int\frac{d^dl}{i(2\pi)^d}\frac{1}{m^2-l^2}\frac{1}{-(l+k_1)^2}\frac{1}{-(l-k_2)^2}\nn
&=& -\frac{1}{16 \pi^2 m^2 \tau}\Big(\mbox{Li}_2(-\tau)+\ln(1+\tau) \ln(-\tau)\Big),\nn
\tau &=& \frac{s}{m^2},\quad s = (k_1+k_2)^2,
\eea
and
\bea
C^{(3)}(s,\mu) &=& \mu^{4-d} \int \frac{d^dl}{i(2 \pi)^d}\frac{1}{(l^2)^2}\frac{1}{(l-p)^2} \nn
&=& \frac{1}{16 \pi^2 s} \bigg\{ -\frac{1}{\e} +\gamma_E
+\ln\left(-\frac{s}{\mufactor}\right)\nn
&&+\e\bigg[\frac{1}{2}\zeta(2)-\frac{1}{2}\bigg(\lns+\gamma_E\bigg)^2
\bigg]\bigg\}\nn
&&+O(\epsilon^2),\nn
s &=& p^2.
\eea 

\setcounter{equation}{0}
\addtocounter{zahler}{1}

\section{Two loop calculation}\label{app:twoloop}

\begin{figure*}
\begin{center}
\begin{tabular*}{\textwidth}{c@{\extracolsep{\fill}}ccccc}
\includegraphics[height=1cm]{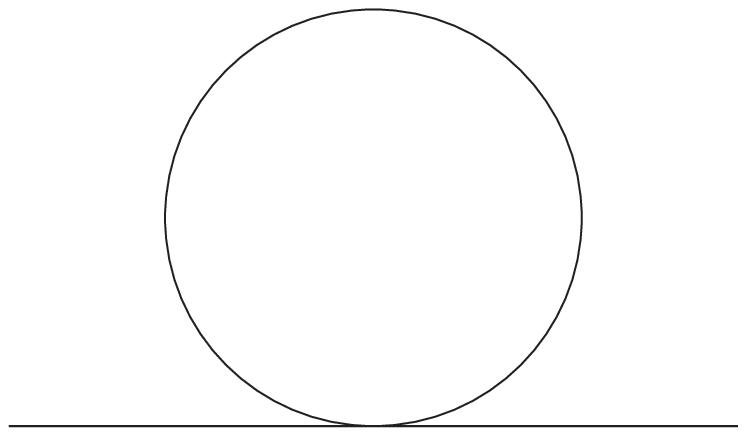}&\includegraphics[height=1cm]{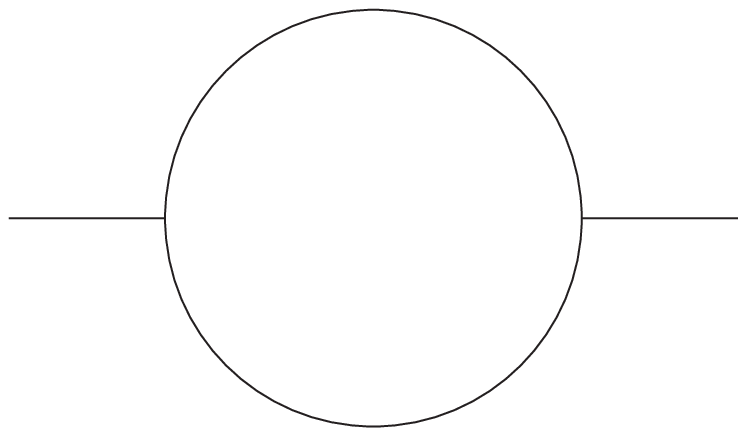}&\includegraphics[height=1cm]{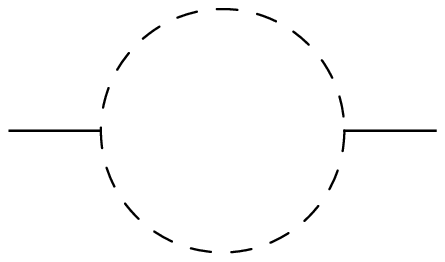}&
\includegraphics[height=1.5cm]{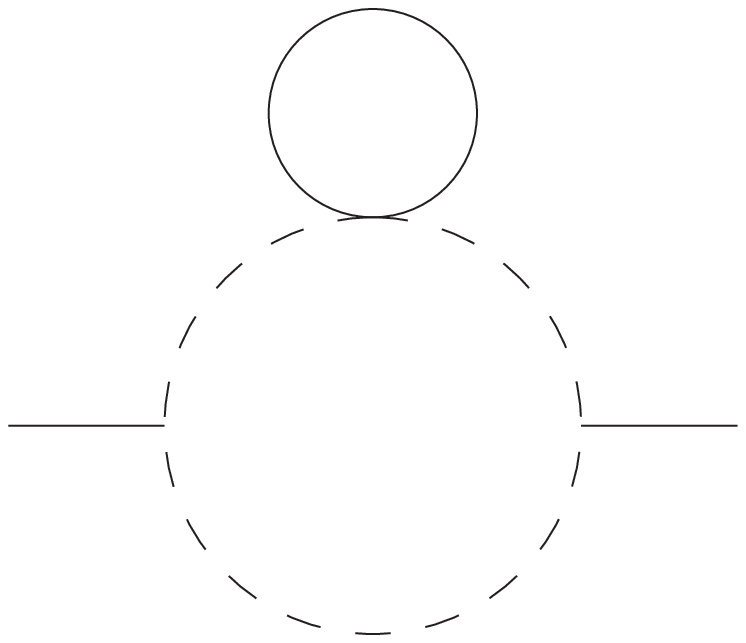}&\includegraphics[height=1.15cm]{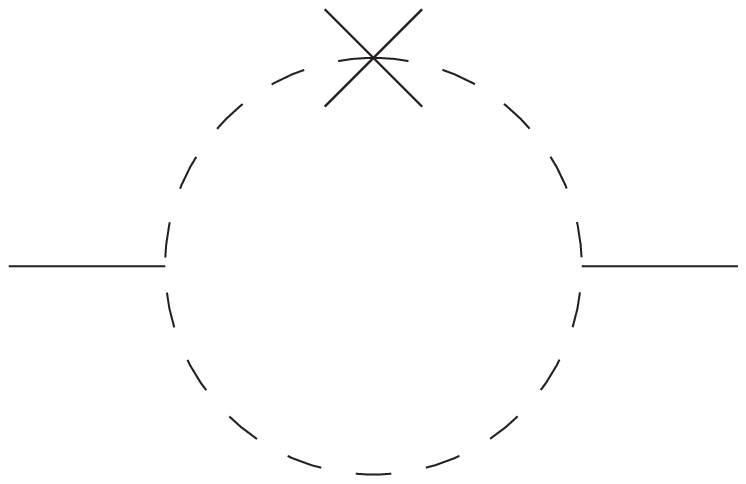}&\includegraphics[height=1cm]{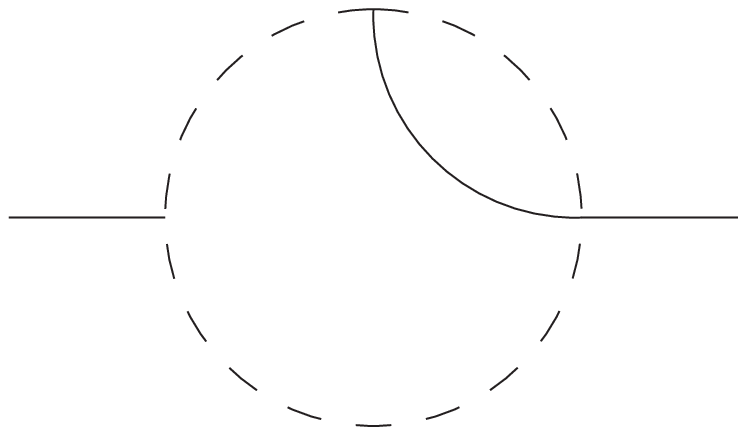}\\
$T_1$&$T_2$&$T_3$&$N_1$&$N_2$&$N_3$\\
\includegraphics[height=1cm]{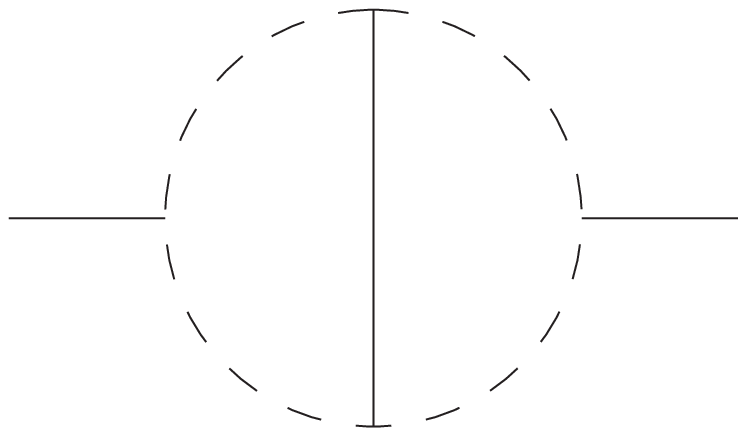}&\includegraphics[height=1cm]{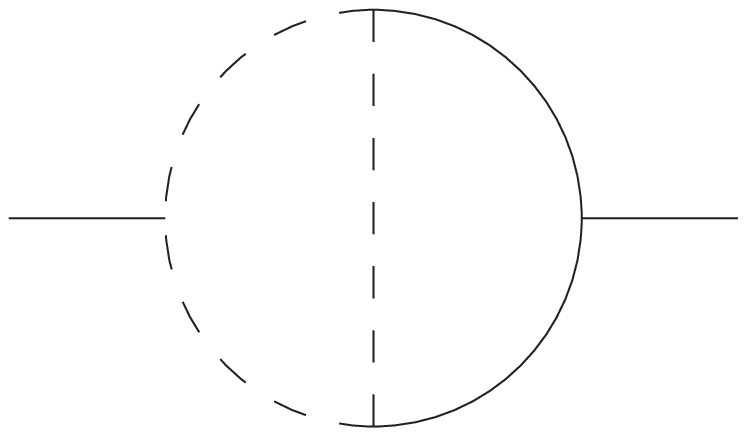}&\includegraphics[height=1cm]{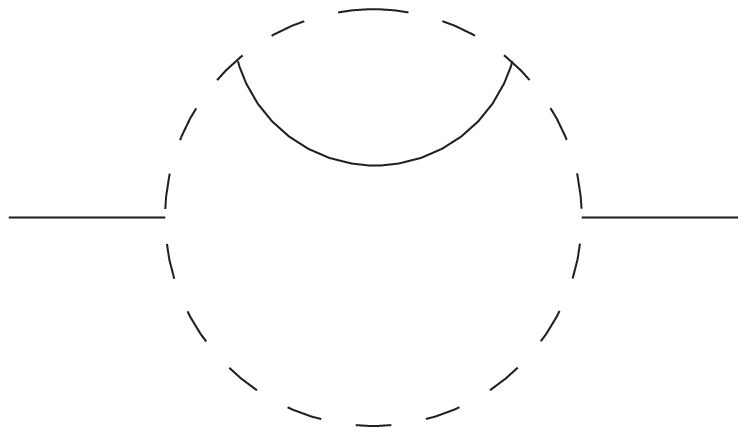}&
\includegraphics[height=1cm]{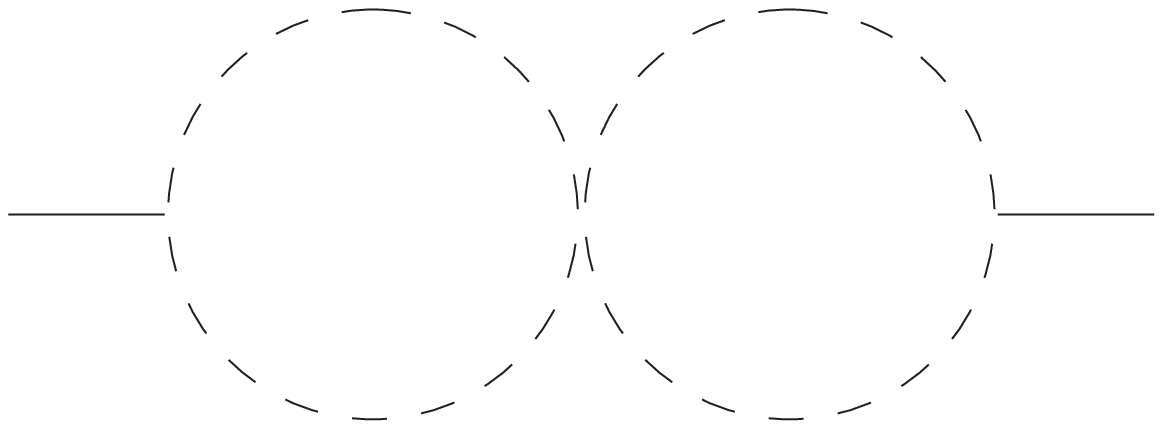}&\includegraphics[height=1.05cm]{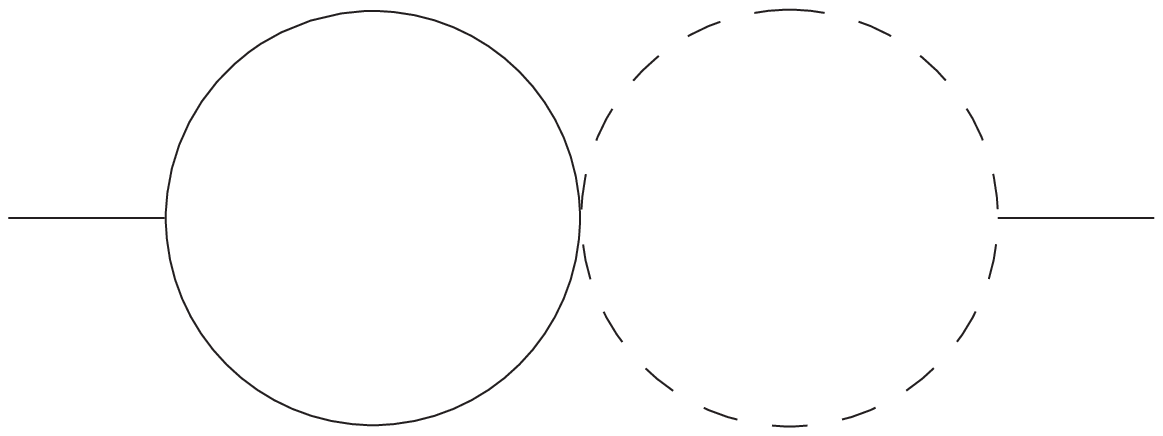}&\\
$N_4$&$N_{5}$&$N_{6}$&$N_{7}$&$N_{8}$&
\end{tabular*}
\caption{Displayed are all the one--loop diagrams and the two--loop diagrams with a branch point at $s =
  0$. The solid line indicates a sigma particle and the dashed line a 
  pion, respectively. \label{fig:2lgraphen}}
\end{center}
\end{figure*}

In the two--loop calculation, we are only interested in the momentum
logarithms. It is therefore sufficient to consider diagrams which develop a
branch point at $s = 0$.
The set of one-- and two--loop selfenergy diagrams which contribute to the discontinuity at threshold are 
shown in Fig.~\ref{fig:2lgraphen}. The analytical 
expressions of the two--loop integrals can be found in \cite{ScharfTausk} and we
adopt the conventions used in this reference. We expand these expressions
around $s = 0$ by keeping the momentum logarithms and expanding the remaining part
in a Laurent series in $s$.
Evaluating 
\beq
G^{(2,\vec{0})}_R(s) = \frac{1}{\Msig^2-s-\Sigma(s)},\label{eq:delta}
\eeq where $\Msig$ is the bare mass of the heavy particle which appears in the
spontaneously broken phase, 
\begin{align}\label{eq:msig}
\Msig^2 &= 2m_r^2\bigg[1-30g_r \lambda-\frac{9 g_r}{16
  \pi^2}\ln\left(\frac{\Msig^2}{\mu^2}\right)+O(g_r^2)\bigg]
\end{align}
yields the result of Eq.~(\ref{eq:2pointcoef}). In Tab.~\ref{tab:contributions} we indicate the contribution of each
diagram $N_x$ to the factor of $L_s^2$
in $G^{(2,\vec{0})}_R(s)$ by inserting only $-i N_x$ instead of the complete
self energies $\Sigma$ in Eq.~(\ref{eq:delta}).
The terms proportional to $1/s$ which stem from the
diagrams containing pion self energy parts as well as the contributions
without an $s$ cancel each other.

\begin{table*}
\caption{Contributions of the different diagrams to the factor of
  $L_s^2$. Every term has to be multiplied with
  $\frac{g_r^2}{1024 m_r^4 \pi^4}$.\label{tab:contributions}}
\renewcommand{\arraystretch}{1.4}  
\renewcommand{\baselinestretch}{1.3}  
\begin{tabular*}{\textwidth}{l@{\extracolsep{\fill}}lllll}
\noalign{\smallskip}
Diagram & Contribution & Diagram & Contribution & Diagram & Contribution\\
\noalign{\smallskip}\hline\noalign{\smallskip}
$ T_1+T_2+T_3 $&$ 90 m_r^2+99 s $ &$N_3  $&$ -24 m_r^2 - 24 s    $&  $  N_{6}  $&$  \frac{24 m_r^4}{s}+24
m_r^2 +18s $   \\
$N_1$ &$\frac{12 m_r^4}{s}+ 12 m_r^2+9s $ & $N_{4}$ & $ 12 m_r^2 + 9s $&
$N_{7}$&$ -60 m_r^2 - 60 s$ \\
$ N_2 $&$ -\frac{36 m_r^4}{s} -36 m_r^2 -27 s$ & $N_5$ & 0& $N_{8} $&$ -18 m_r^2 - 2s$  \\
\noalign{\smallskip}\hline
\end{tabular*}
\end{table*}

\setcounter{equation}{0}
\addtocounter{zahler}{1}

\section{Dispersive Calculation}\label{app:dispersive}

\begin{figure}[h]
\centering
\begin{tabular}{ccc}
\includegraphics[height=1cm]{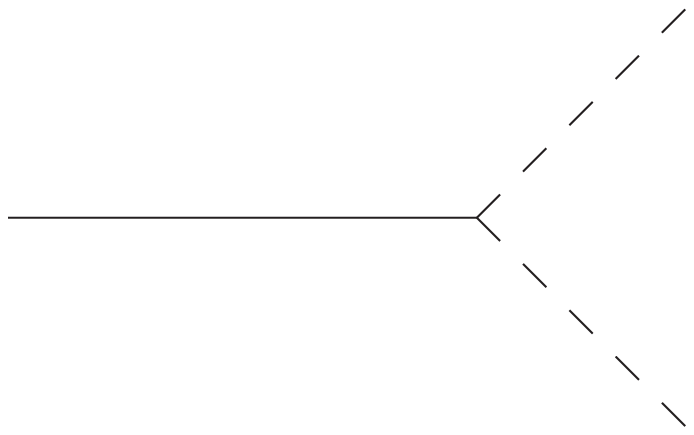}&\includegraphics[height=1cm]{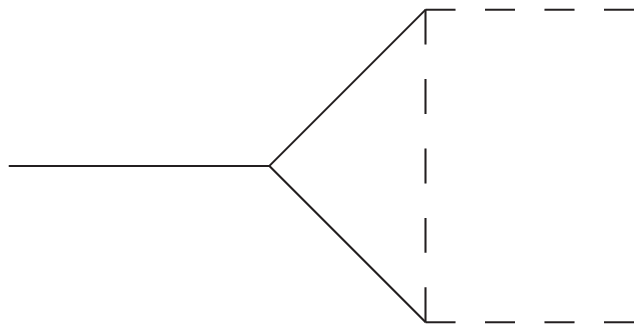}&\includegraphics[height=1cm]{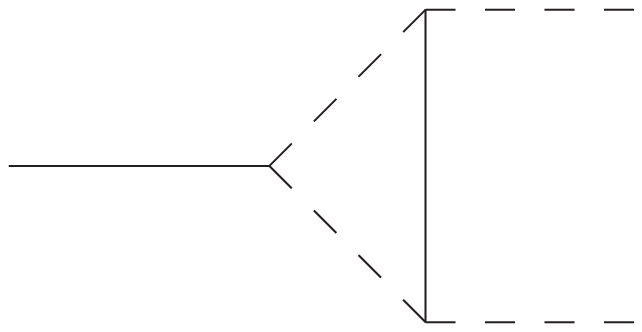}\\
$G_0$&$G_1$&$G_2$\\
\includegraphics[height=1cm]{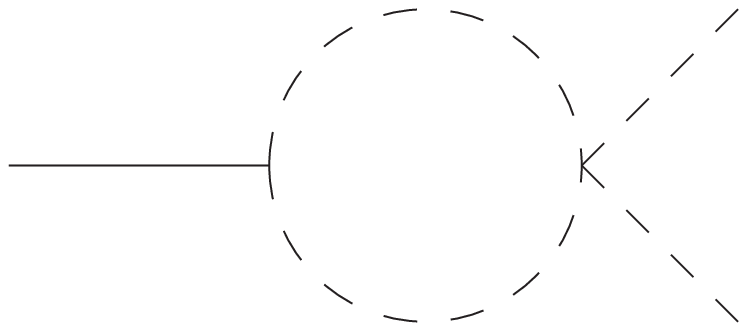}&\includegraphics[height=1cm]{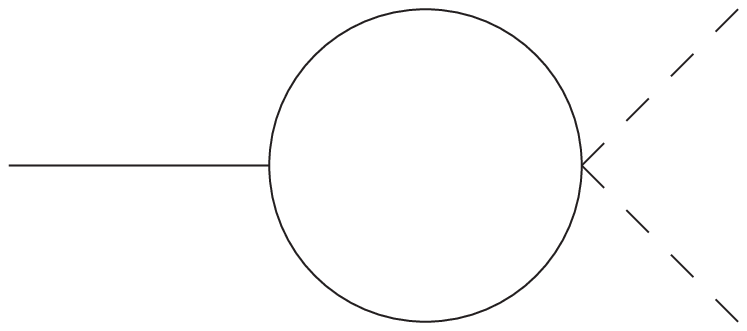}&\includegraphics[height=1cm]{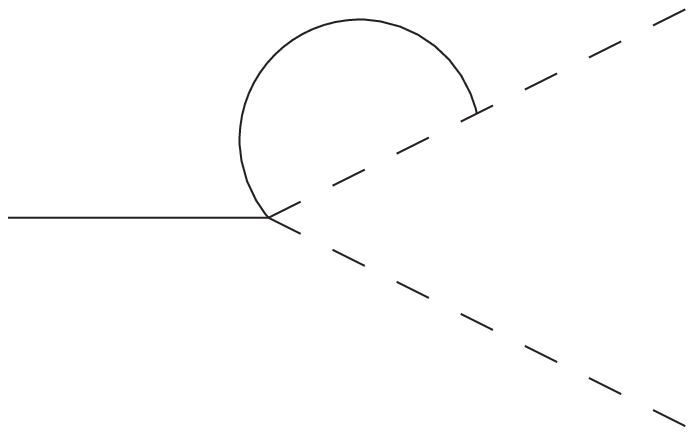}\\
$G_3$&$G_4$&$G_5$\\
\end{tabular}
\caption{Diagrams contributing to the matrix element $\langle 0 | \phi(0) | \pi^a
  \pi^b \rangle $. The solid line denotes a sigma particle and the dashed
  line stands for a massless pion, respectively.\label{fig:dispgraphen}}
\end{figure}

To calculate the discontinuity at two loops, the 1PI
truncated diagrams shown in Fig.~\ref{fig:dispgraphen} and the one--loop
diagrams indicated in Fig.~\ref{fig:2lgraphen} are
required. We only need the analytical expressions of the diagrams for $s$
small compared to the mass of the sigma particle.
Performing the phase space integration
\bea
\rho(q^2) &=& (2\pi)^{3}\frac{1}{2}\sum_{a,b} \int d\mu(k_1)d\mu(k_2)
\delta^{(4)}(p-k_1-k_2)\nn
&&\times \abs{\mael{0}{\phi(0)}{\pi^a(k_1)\pi^b(k_2)}}^2,
\eea
where $d\mu(k)$ is the Lorentz invariant measure
\beq
d\mu(k) = \frac{d^3k}{(2\pi)^3 2k^0},
\eeq
leads to the discontinuity 
\bea\label{eq:rhosm}
 \rho(s)
 &=& \frac{1}{16 \pi^2} \Bigg[\left(\frac{3 }{2 m_r^2} +O(s)\right)g_r
   +\bigg(-\frac{3}{16 \pi^2 m_r^2}\nn
&&+\frac{9}{16 \pi^2 m_r^2}\ln\left(\frac{2m_r^2}{\mu^2} \right) +O(s) 
     \bigg)g_r^2\nn
 && + \left(-\frac{3 s
     }{16 \pi^2 m_r^4}+ O(s^2) \right)\ln\left(\frac{s}{\mu^2}\right) g_r^2\nn
&& + O(g_r^3) \Bigg],
 \eea
which agrees exactly with the discontinuity calculated directly from our two--loop result.\newline
Evaluating the phase space integration in $d$ dimensions, we checked the
discontinuities of the single two--loop diagram $N_{3}$, $N_{4}$,
$N_{5}$, $N_{7}$ and $N_{8}$.

\setcounter{equation}{0}
\addtocounter{zahler}{1}

\section{Summation of scale dependent leading logarithms} \label{app:summation}

Splitting up the mass and momentum logarithms as described in the text,
it is possible to sum the logarithms $L_{\mu}$ to all orders. As described in
section \ref{sec:RG}, we collect all leading logarithmic terms which exhibit a factor
of $(s/m_r^2)^t$ in the function $f_t(x)$,
\begin{align}
f(x) &= \sum_{t=0}^{\infty}\left(\frac{s}{m_r^2}\right)^tf_t(x), &f_t(x) &=
\sum_{k=0}^{\infty} d^{(k,t)}_k x^k,\nn
x &= g_r L_{\mu}. 
\end{align}
Solving the corresponding differential equation with the initial condition
$f_t(0) = 1/2^t$ one obtains
\bea
f_t(x) &=& \frac{1}{2^t}\bigg(1-\frac{3}{4 \pi^2}x  \bigg)^{\frac{1}{2}(1+t)}.
\eea
The series in $s/m_r^2$ can also be summed and yields
\begin{align}
G^{(2,\vec{0})}_R(s) &= \frac{1}{2 m_r^2\left(1-\frac{3}{4\pi^2}x 
\right)^{-\frac{1}{2}}-s}+\cdots, \label{eq:modprop}
\end{align}
where the ellipsis denotes all the subleading terms. Choosing $\rho^2=2m_r^2$ and calculating the real part of the
zero of the modified inverse propagator Eq.~(\ref{eq:modprop}) one recovers
the result from the subsection \ref{sec:zeroinv}.\newline 
To establish connection
to the recursion relations derived in section \ref{sec:RG}, we express the coefficients
$d$ by means of the coefficients $a$,
\beq
d^{(N,t)}_N = a_{N,0}^{(N,t)}+a_{N-1,1}^{(N,t)}+\cdots+a_{0,N}^{(N,t)}.
\eeq
Therefore, the function $f(x)$ includes the coefficients of the leading
momentum logarithms. However, as we have seen in section \ref{sec:RG}, the RGE
do not allow to sum them separately.
\end{appendix}

\end{document}